\documentclass[a4paper,twoside]{article}

\usepackage{epsfig}
\usepackage{subcaption}
\usepackage{calc}
\usepackage{amssymb}
\usepackage{amstext}
\usepackage{amsmath}
\usepackage{amsthm}
\usepackage{multicol}
\usepackage{apalike}
\usepackage{SCITEPRESS}     % Please add other packages that you may need BEFORE the SCITEPRESS.sty package.

\begin{document}

\title{Investigate how developers and managers view security design in software}

\author{\authorname{Asif Imran}
\affiliation{\sup{1}Department of Computer Science, \\ California State University San Marcos, \\ 333 S Twin Oaks Valley Rd, San Marcos, CA 92096, USA}
\email{\{aimran\}@csusm.edu}
}


\abstract{Software security requirements have been traditionally considered as a non-functional attribute of the software. However, as more software started to provide services online, existing mechanisms of using firewalls and other hardware to secure software have lost their applicability. At the same time, under the current world circumstances, the increase of cyber-attacks on software is ever increasing. As a result, it is important to consider the security requirements of software during its design. To design security in the software, it is important to obtain the views of the developers and managers of the software. Also, it is important to evaluate if their viewpoints match or differ regarding the security. Conducting this communication through a specific model will enable the developers and managers to eliminate any doubts on security design and adopt an effective strategy to build security into the software. In this paper, we analyzed the viewpoints of developers and managers regarding their views on security design. We interviewed a team of 7 developers and 2 managers, who worked in two teams to build a real-life software product that was recently compromised by a cyber-attack. We obtained their views on the reasons for the successful attack by the malware and took their recommendations on the important aspects to consider regarding security. Based on their feedback, we coded their open-ended responses into 4 codes, which we recommended using for other real-life software as well.}
\maketitle
%\onecolumn \maketitle \normalsize \setcounter{footnote}{0} \vfill

\section{\uppercase{Introduction}}
\label{sec:introduction}

The recent trend in world events has made cyber threats an issue of ever-increasing importance. Traditionally, software engineers have not focused on building security into their systems. The main concern has been to make data and networks secure through peripheral security measures. Under the current circumstances, software companies are striving to include quality attributes such as security in the software design and analysis phase. The concept of \textit{"building security into the software"} is a talked about topic in the industry, however, we do not see any actionable steps from the companies on how to make their existing and new systems adhere to the security requirements.

Sparked by the requirements of the industry, recent studies have shown an increase in cyber threats for real-life software \cite{ulsch2014cyber}. Researchers identify defensive mechanisms at the hardware and operating systems level to counter malware threats \cite{lawson2019cyber}. Software industries which produce in-house software have focused on deploying firewalls, whereas increasing number of software services are being provided online using the internet \cite{kumar2008locking}. Security needs to be ensured for real-life software at the design and development stages to make them cyber-attack proof \cite{beznosov2008security}. Despite the requirement to make the software more secure, what current challenges prevent software engineers from building security into their systems need to be explored in greater detail. 

In this paper, we described our approach and findings based on our investigation of how software teams looked into the current challenges which prevented the security of software applications. 
 A survey participated by 9 respondents of a software company who were responsible for designing software that recently got compromised by a malware attack was conducted, and we aimed to determine what factors which prevented the secured design of the software. The 9 respondents were divided into two teams, who worked on different components of the compromised project. For purpose of anonymity, we cannot publish the names of the respondents, product or the company. We tried to identify if the developers and the managers shared the same views related to software security challenges, or do their views differed. We further explored what recommendations in terms of collaboration and design challenges suggested by developers and managers which would enable mitigation of the current challenges. 

Under the current circumstances, it is important to identify the viewpoints of software developers and managers regarding security vulnerabilities in software. Motivated by the research in \cite{storey2022developers}, we explored the perspectives of developers and managers on the reasons for malware vulnerability in software. 

The survey of managers and developers highlighted that managers and developers agreed on certain criteria of security, however, they differed in their views on other criteria. We explored why being in the same cohort they had differing views on certain aspects of building security into their systems. Our findings showed that both developers and managers agreed that software can become vulnerable to malware attacks due to the \textit{lack of skills} of the project team members who designed it, \textit{negligence or ignorance} by end users, and \textit{trade-off} between security and key aspects. 

Developers and managers could not reach an agreement if \textit{work environment and well-being} of the project team had a role to play in getting the team more motivated and designing security into the software systems. It was seen that although many developers stated the role of team cohesiveness towards designing secure software, managers emphasized more on building skills of developers.   

The major contributions of this paper include the following:

\begin{itemize}
    \item Survey of software developers and managers who worked in two groups to design a commercial software that was recently compromised by malware.
    \item An empirical evaluation of the common and differing views of  developers and managers on what caused the malware attack to be successful on their software.
\end{itemize}

The rest of the paper is organized as follows: Section \ref{methodology} explains the methodology of the research. Section \ref{results} presents the survey design and selected responses. Section \ref{surveyfindingsanalysis} highlights the analysis and findings of the survey and states important recommendations regarding developers and manager views of software security design. Section \ref{relatedwork} discusses the related work in this area, Section \ref{threats} identifies the threats to validity of this research, and Section \ref{conclusion} concludes the paper.

\section{Methodology and Survey Setup}
\label{methodology}

To determine the views of developers and managers on the cause of security vulnerabilities in software, we surveyed professionals from a large software company who worked on a product that recently suffered from cyber attacks. Our research questions were as follows:

\begin{enumerate}
    \item How developers and managers define reasons why software suffer from malware attacks?
    \item Are developer’s view of security threats to a software aligned with each other and views of managers?
\end{enumerate} 

Our goal is to identify the views of developers and managers regarding software security. We try to find the recommendations of the developers and managers regarding mitigating security vulnerabilities in software design based on their real-life experience.

\section{Survey design}
\label{results}
To ensure that the survey questions were accurate, we worked closely with two developers and one manager who helped us to recursively design the questionnaire. We conducted the survey recursively and the team of software professionals helped us pilot it. 

The survey was made available to the respondents in March 2022. Based on our prior association with the software company, we were able to contact the respondents via email. We contacted the developers and managers of two small teams within a large company, who combinedly designed and developed one product that faced cyber attacks on January 2022 and was compromised. 
\begin{figure*}
    \centering
    \includegraphics[width=14cm, height=6cm]{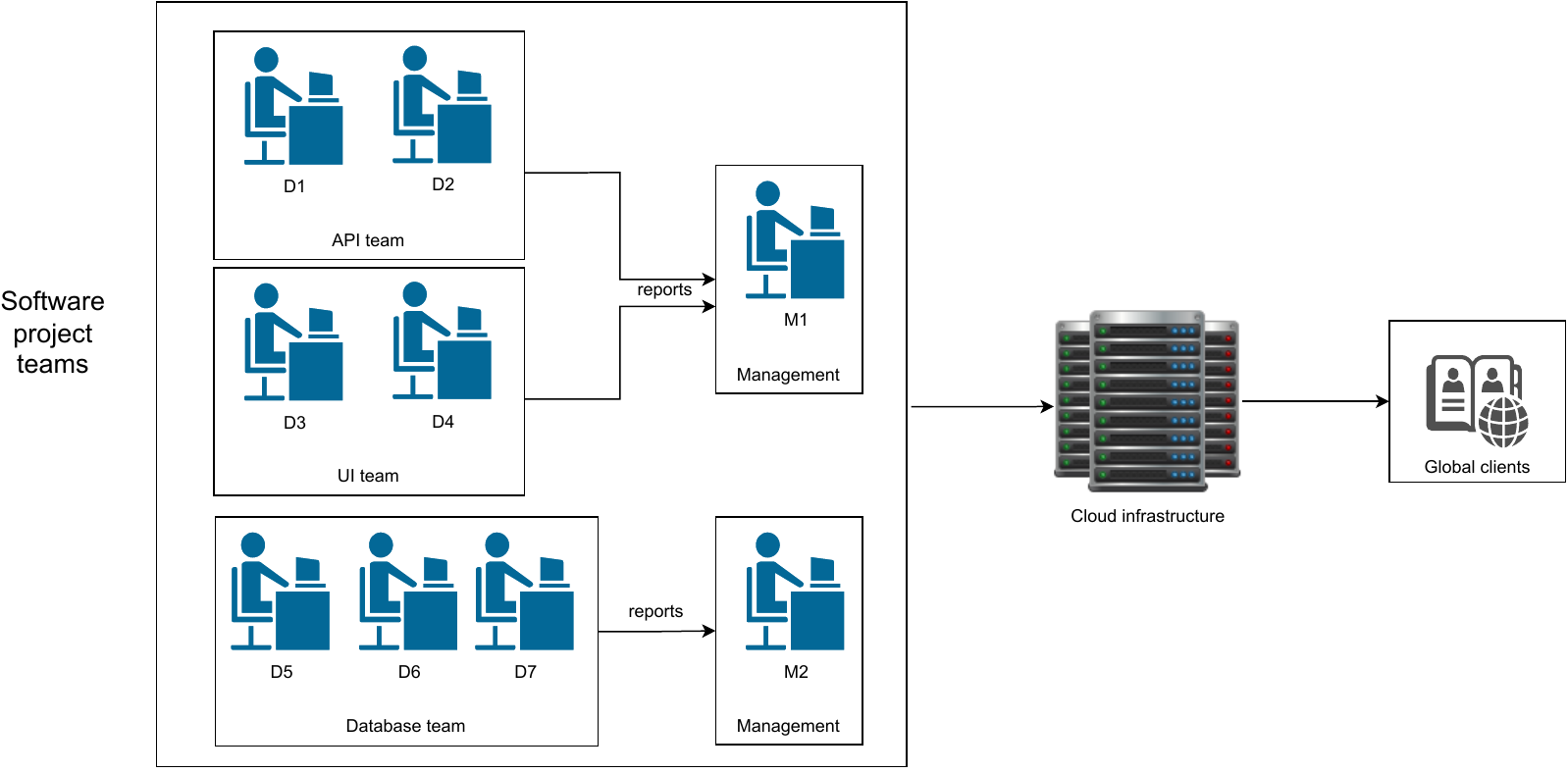}
    \caption{Project team structure for the surveyed software.}
    \label{team}
\end{figure*}

We advertised the survey to them as a general survey to obtain their views and recommendations for software security design. The survey was kept available for 20 days, and 2 managers and 7 developers responded. Our initial results showed that both developers and their managers had a clear concept of how each perceived software security in terms of user behavior, skills, and trade-off. However, they did not seem to agree on the perspective of integrating security into software design and its relationship to the work environment of the team.

\subsection{Survey structure}
We identified the respondents based on our prior association with the company. The respondents worked in the same software team that designed and implemented an \textit{application interface (api)} that was used by the core product of the company which was a \textit{Database Management System (DBMS)}. 

The \textit{api} acted as a middleware used by the DBMS to launch itself in the Windows platform. The team structure and roles of the 9 respondents were highlighted in Figure \ref{team}. This was a large software company, that served many energy and storage industries in the USA and abroad. The surveyed company and respondents have been kept anonymous for the purpose of security. We surveyed 2 managers who were leading the team which designed and developed the \textit{api} that acted as a middleware between DB and OS. We surveyed 7 developers who were working under the managers, we tried to gather their views on software security design of the compromised \textit{api}. 

Based on our earlier relationship with the company we learned that the \textit{api} that was designed had been compromised by \textit{Azorult} malware that lurked in the host operating system of a client, thereby it successfully obtained access to the dashboard of the DBMS. \textit{Azorult} malware was used by the attackers to infect the target software. This malware was first discovered in 2017 \cite{rendell2019understanding}. 

\textit{Azorult} is primarily used to confiscate user information on communication platforms like Discord. It is also used by the attacker to steal the browser history, cookies, and cryptocurrency keys of the victims. Finally, the data is sent back to the attacker via a zipped file. Recently, malware has been found that \textit{Azorult} can load other malware and the recent versions have improved delivery mechanisms compared to earlier ones.

Prior to data collection, we consulted industry experts in survey design and asked them to answer the survey as a test pilot. The survey included 16 questions that were conducted in a semi-structured format. We iteratively tried to determine developers and manager views. The responses were coded by the second and third authors and assigned to the the pre-specified codes. Any ties in coding were dissolved by the author. It took 9.10 minutes on average to complete which was an acceptable completion time, thus not consuming the excess time of the respondents. 

The respondents included both developers and the manager of the \textit{api} project. We aimed to determine the views of the developers and managers regarding any issues related to software engineering practices that led to a faulty design thus resulting in the malware compromise. We aimed to identify if the views of developers and managers differed in this regard and tried to determine \textit{"why"} the difference existed. 

\begin{figure*}
    \centering
    \includegraphics[width=14cm, height=5cm]{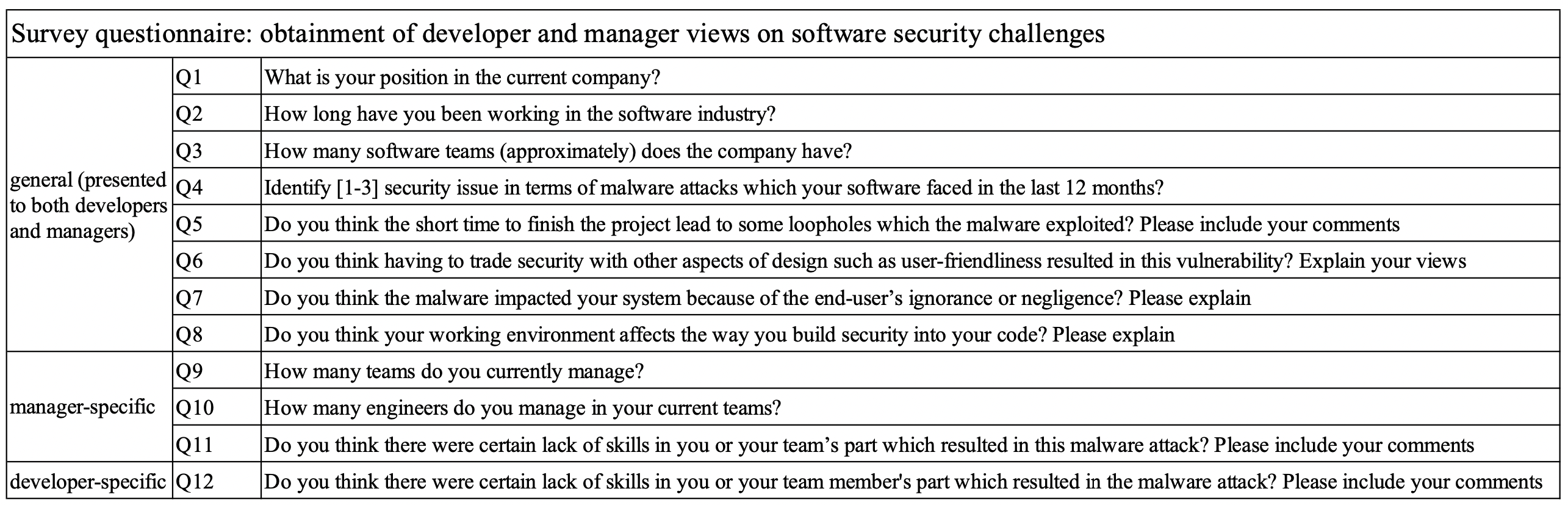}
    \caption{Survey instrument to obtain viewpoints of developers and manager on software security challenges.}
    \label{questionnaire}
\end{figure*}

Additionally, any recommendations by the respondents based on their experience during that process on how to securely design similar software for DBMS to thwart such attacks were discussed here. The survey instrument is shown in Figure \ref{questionnaire}. The questionnaire precisely asked the respondents to relate the cause of the detected cyber attack compromise to the reasons like \textit{skills of the development team, user behavior, trade-off, and work environment}. The respondents provided their responses which show-cased their different viewpoints. We recorded all the responses and mapped those to the software attributes described above. The data were compiled and presented in the following section.

\subsection{Survey takeaways}

We found that developers and managers agreed to some extent regarding the cause of malware attacks, however, they seemed not to agree on the steps which were required in software design to make sure the malware cannot attack in the first place. The open-ended responses were coded by us in an inductive manner, and initial codes namely \textit{Skills (S), User behavior (U), Trade-off (T), and Work environment (E)} \textit{(SUTE)} were refined after further discussion. Following are the codes for developers and the manager. Regarding skills, the following managerial feedback can be considered.

The two managers have been identified as \textit{M1} and \textit{M2} respectively, and the seven developers are identified as \textit{D1, D2,...,D7} respectively. Regarding the code \textit{S}, manager \textit{M1} stated: \textit{"yes, I personally think virus writers are better programmers than some of my team members, the issue is when we make a team, we select people who are good software engineers, however, to address security we should have a policy where all teams need cyber security experts"}. The response provided by \textit{M2} is similar and notable, \textit{"we design and develop large database apps that communicate with the host OS. Now, I have a big team of engineers working on our project and someone is always a weak link"}. 

Regarding skills and expertise, developer \textit{D1} stated that \textit{"Malware developers continuously develop new ways to get around design to which is deployed to prevent malware. On the other hand, our skills are mostly applicable for after-attack situations}. Another developer, \textit{D2} stated that \textit{"we react to attacks, rather than focusing on predicting and preventing the malware attack in the first place"}. \textit{D3} mentioned \textit{"our teams only constituted of domain experts and software developers, we definitely lacked the inputs of a cyber security expert, this is something we are planning to include in our future projects"}. Another developer, \textit{D6} noted that \textit{"I primarily focused on finishing the different functional requirements of the api, which included incorporating the different features, never really got time to check if the features meet security guidelines"}.

Regarding user behavior, \textit{M1} stated, \textit{"Most malware got into our software through the OS because of the user's mistakes like clicking on suspicious links that allowed the Azorult to infect the system and ultimately gain access to our api}. \textit{M2} mentioned that, \textit{Users are “click robots”, they are asked to click something, and they click it"}. From the developer perspective, we got feedback from \textit{D5} such as \textit{"users do not keep their systems up to date, because they are not IT people, they do not have IT people"}. 

Also, another developer, \textit{D7} responded by stating, \textit{"They open attachments from people they do not know, or from people who pretend like someone they know"}. Developer \textit{D1} mentioned \textit{"The weakest vulnerability of our api was probably the human element. User behavior is the chief reason for creation of access points which were exploited by the Azorult"}. \textit{D2} stated that \textit{"I think besides developers and managers, general users with minimal technical knowledge should be trained about system health, awareness about suspicious behaviors, and creating strong passwords"}.

Tradeoff is another important consideration for security. According to manager \textit{M2}, \textit{"inevitably some decisions are made in favor of usability, rather than security, so there was a tradeoff between usability and security}. The reason for the tradeoff was mentioned as \textit{otherwise people will not buy and use our solution"}. Developers also highlighted that quality requirements such as security was traded off for this project. 

Developer \textit{D4} mentioned that \textit{"it is a cat and mouse game, think of building a house, why cannot architects and contractors build homes and business buildings that are impervious to burglars"}? \textit{D3} responded \textit{"amount of time I can spend doing actual development compared to attending team meetings, I had to tradeoff my productivity and time with attending meetings, some of which were not really useful"}. 

Developer \textit{D2} mentioned that \textit{"we wanted to ensure least privileges in all aspects of the software, however, user-friendliness often came in the way and we had to make the software more easily accessible by the users, which was definitely a wrong decision in my opinion"}. It was visible that tradeoff happened due to usability, ensuring least privileges, and attending team meetings in a large institution not directly related to the project.

Regarding environmental condition and personal well-being, manager \textit{M1} noted, \textit{"personally I think work environment does not impact security aspects of the software rather the focus is more on building developer skills"}. Another response from manager \textit{M2} was such that \textit{"developers tend to focus on how many artifacts they produce in a sprint whereas I tell them to get the job done efficiently and preserve highest quality, we always do that in good team spirit so I think team environment is not a concern"}. It can be inferred that both the managers believe that their teams have good working environments and there is no issue in this regard. We analyze the responses obtained from the developers in the following.

The developers had a different opinion regarding the personal well-being and work environment and their impact on including quality attributes in software design. Feedback from developer \textit{D1} included \textit{"When I feel engaged in the work I am doing, I do not feel held back and try to implement new security designs in the component I was developing"}. Additionally, we obtained feedback from developer \textit{D2} such as \textit{"when I am truly engrossed in a task, I felt that I worried more about non-functional requirements like security}. 

Developer \textit{D3} stated, \textit{"My manager seemed to avoid many questions I had regarding the api during the design and development stages, after some time, I lost a lot of motivation and tried my best to obtain the answers to my questions using the web. If my manager provided timely feedbacks and appreciated my efforts, I think we would have worked in a more cohesive environment"}. \textit{D4} mentioned that \textit{"I personally was not impressed on how our team communication was maintained, we did not get feedback in a timely manner from other members, including developers and managers, so it was a 9 to 5 job where I was concerned with finishing my tasks to meet functional requirements only"}. Finally \textit{D7} said, \textit{"I focus on building real software and gaining new knowledge, and this can only happen when I am working in a place where people care and provide timely answers to my queries"}. As a result, multiple respondents mentioned the requirement of a supportive team to build secured software. 

\begin{table*}
\centering
\begin{tabular}{ |p{3cm}| p{11cm} |  } 
 \hline
 \textbf{Code} & \textbf{Description of code}  \\ \hline 
 S: skills & do not possess the skills required to remove all bugs which may result in security vulnerabilities \\
 \hline 
 U: end user behavior & ignorant and negligent activities from end users result in malware attacks succeeding\\
 \hline 
 T: trade-off & trade-off between completion of user stories to security\\
 \hline 
E: environment & impact of personal well-being and team cohesiveness towards designing a more secure software \\
 \hline 
\end{tabular}
\caption{Description of the codes.}
\label{codessurvey}
\end{table*}

\section{Analysis of findings}
\label{surveyfindingsanalysis}

The thematic analysis was conducted by one of the authors of the paper based on the responses submitted by survey participants during the interview. The survey questionnaire was prepared in an online form that was provided to the respondents. The data was exported in an excel file and the responses were recorded that were used to review and complete the findings. The exporting of the data into an excel file helped to harness the power of the survey tool that we used and it helped to leverage the design of our survey for analysis. Responses to each question were analyzed on a per-question method, which resulted in finding similarities across all responses and coding the results. The codes were provided in Table \ref{codessurvey}. 

The survey had open-ended questions as seen in Figure \ref{questionnaire}. It was important to code the open-ended question to obtain correct takeaways from those. We proceeded to code the open-ended responses recursively and inductively. Two authors initially coded all the open-ended responses. Next, we conducted a dedicated session where all three authors analyzed the coded responses and reached an agreement. After multiple rounds of coding, 4 codes were determined to be applicable to analyze both developer and manager views. We proceeded to analyze the coded responses to determine any similarities or differences in the perspectives of developers and managers.

The four main codes that resulted from coding the productivity definition question are as follows: skills of the developers and managers, user behavior when using the software, trade-off between security and usability, and developer satisfaction and work environment. We noted that these four codes from recursive coding practice are in-line with the dimensions provided by the SPACE framework \cite{forsgren2021space}.

\begin{figure}
    \centering
    \includegraphics[width=7cm, height=6cm]{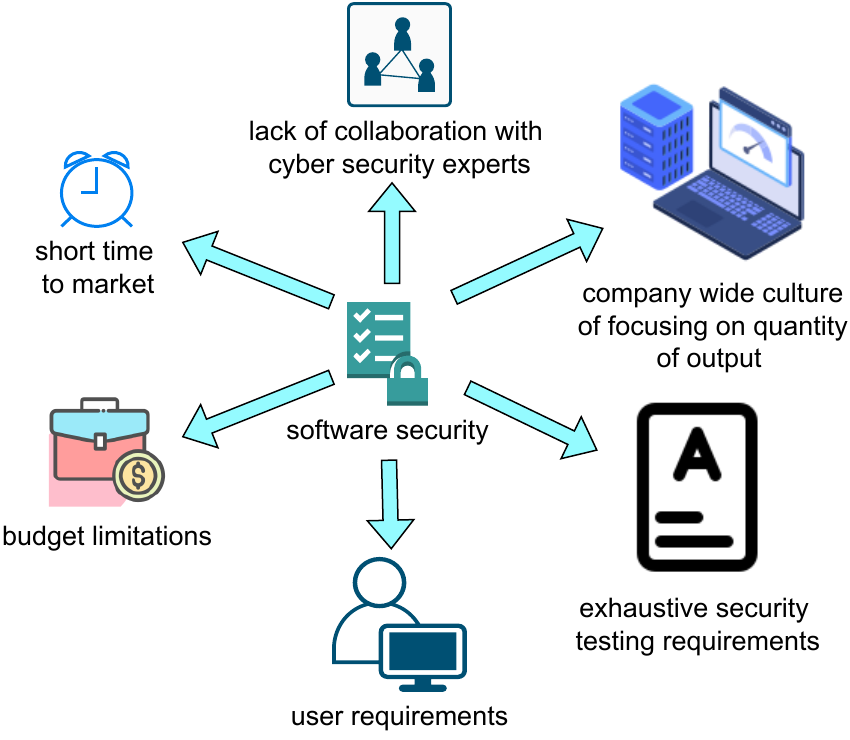}
    \caption{Identification of the different reasons according to developers and managers which lead to trading off security in software design.}
    \label{tradeoffsecurity}
\end{figure}

\begin{itemize}
    \item \textit{S}: both the developers and managers agree that there is a lacking in the current skill set of engineers and recommended new upcoming engineers to \textit{“think out of the box and focus on developing malware threat prediction-based action skills”} rather than \textit{“reactive skills”}.  In this regard, the software teams should include cyber security experts in them.
    \item \textit{U}: 100\% of the respondents stated that user negligence has a role to play in software getting impacted by malware. One thing to remember is that malware primarily does not attack the software, it attacks OS first in most cases. How users use the OS is very important in this regard.
    
    \item \textit{T}: over 78\% of respondents identified trading security for multiple reasons that are highlighted in Figure \ref{tradeoffsecurity}. \textit{Limited budget, exhaustive testing of security features, lack of collaboration with cyber security experts, strict time to market, company culture,} and \textit{user requirements} are a few attributes that forces the team to tradeoff security for user-friendliness. Around 22\% of respondents stated that they will not tradeoff security for usability as they stated \textit{"software which follows secured design with the highest importance is likely to achieve higher user satisfaction in the long run than software that looks to reduce security for usability"}. The respondents also mentioned the importance of collaboration between development team and cyber security experts to achieve highly secured software. 
    
    \item \textit{E}: although developers and managers agree on the S, U, and T aspects of security design into their systems, they have different views when it comes to work environment. Developers addressed the relationship of team cohesive, caring work environment and building security into their systems. Developers defined \textit{“building secured software”} as related to their work environment satisfaction whereas managers focused more on improving the skillset of developers to achieve this goal. 
    
    Developers and managers addressed software security challenges in terms of \textit{SUT} codes, however, they seem to differ in their viewpoints regarding a cohesive work atmosphere. More specifically, developers and managers, as cohorts, were not in-line with their viewpoints regarding the impacts of a cohesive work environment and designing security into the software. Managers tend to view security design as dependent on the \textit{SUT} codes rather than the \textit{E} code. On the other hand, developers, emphasized the need for \textit{E} code, which is a friendly work environment, as a critical aspect if addressing security in software design.
\end{itemize}

\section{Related Work}
\label{relatedwork}
The importance of methods, processes, and tools for improving the security of software created by a wide range of developers is emphasized in existing research \cite{beznosov2008security}. Modern software works mostly in a connected and collaborative manner, therefore they need flawless access to other systems in different geographic locations \cite{imran2013provintsec} \cite{imran2016web}. Security needs of modern-day software need to be addressed from the perspective of software design, developer skills, user behavior, and legislation governing software engineering \cite{kolodziej2011meeting}. At the same time, software engineers need to collect security requirements during the software requirements elicitation phase.

Security is mostly considered an afterthought and incorporated only after the software has been developed \cite{4591701}. It was stated that there is a lack of processes to quantify the security of a \textit{Software Development Life Cycle (SDLC)} artifact. As a result, the number of vulnerabilities was detected to calculate a vulnerability index of an \textit{SDLC} artifact which acted as an indicator of the current vulnerabilities. However, a mechanism to code the different areas of human factors which lead to the vulnerabilities in design is required. 

Activities to solve software security issues are mainly considered after completion of the coding phase \cite{jurjens2005secure}. Security professionals address security concerns by using tools like antivirus, proxies, firewalls, and intrusion prevention systems mainly during the final stages of \textit{SDLC} \cite{jurjens2005secure}. However, addressing security challenges after the coding phase may require more expensive reworks \cite{4591701}. As a result, security needs to be considered at an early stage of \textit{SDLC} through the incorporation of feedback from the project team. 

Existing research focuses on the security of software primarily designed for military, government, and safety-critical systems where budget and time to market have less priority and security is the primary concern \cite{bressan2021variability}. However, the focus should be also provided to software which is designed by software industries where cost, time to market, and usability are the main concerns, and security is considered a non-functional requirement.

Software security in earlier publications stated the importance of security practices such as access control, firewalls, encryption, and authentication \cite{g2021achieving}. We agree that all security features are important, however, from a software engineering perspective, the processes and designs followed to build the software are equally important. Therefore, we think further research should be conducted in this regard.

The importance of secured software design was highlighted in \cite{zhang2022example}. The authors mentioned that following a structured security design practice during software development should be given importance. They argued that although redundancy can help during random cyber attacks, focusing on reliability only will not ensure security against well-planned and coordinated cyber-attacks since such attackers are capable of compromising redundant copies of the same system if they can compromise the deployed version. 

The importance of collecting security requirements besides regular software requirements is emphasized in \cite{tondel2008security}. The authors stated that correct identification of security requirements will make sure that the software performs the correct operation. Whereas, wrong security requirements will lead to failure of performance by the software, which often leads to finger-pointing by the engineers. Also, if the software team lacks any security experts, then these problems will increase manifold. In addition to collecting software requirements, the process and time when to collect these requirements should be specified after discussion with developers and managers. 

Threat analysis has been proposed as a critical activity to identify and prioritize different threats during software design and development \cite{ingalsbe2008threat}. Given the situation where cost, usability, and time to market take greater importance to security, an effective threat analysis mechanism will enable software professionals to detect the most critical threats and resolve those with the given budget and time. However, besides threat analysis, importance should be provided to gathering real-life experiences of both developers and managers to prioritize threats in future software.

\section{Threats to validity}
\label{threats}

We identify the common threats to validity as experienced by similar qualitative research. Generalization of the survey results beyond the scope of the sampled respondents needs to be conducted with care. We interviewed only industry experts and not the clients of the software, also we interviewed the teams related to designing and developing the software, the opinion may vary among teams working on different software projects in the same company. We limited our scope to one software company only, however, the findings may not be reflected across different sized companies at different geographic locations. Furthermore, our number of respondents was small, and hence our results may not be generalized to other software teams. More specifically, the number of managers we interviewed was low.

\section{Conclusion}
\label{conclusion}
We determined that developers and managers had different viewpoints regarding the work environment and its impact on designing security into software. Based on the findings regarding the misalignments between the views of developers and managers about quality, we suggested that the development team and managers should regularly communicate their views on security design and state what the work environment means to them. We suggest, even before obtaining security requirements and starting a security-focused software design process, developers and managers should first communicate what security level they want the software to achieve. At the same time, they should openly discuss the necessary skills, tools, and work environment which is required to build security into the software.

We also recommend software companies use our provided coding mechanism called \textit{SUTE} to evaluate the security challenges faced by their software teams. Although the codes have been tested on two teams in one company, we believe they should be piloted across different projects to establish our recommendations. Our coding mechanism reveals a misalignment between developers and managers when it comes to a working environment and security design. Similar findings have been stated in earlier research efforts \cite{francca2018motivation}, \cite{graziotin2013happy}. Hence, we believe the proposed coding technique can help companies evaluate security challenges in existing projects. 

At the same time, our coding mechanism can be used with the popular \textit{SPACE} framework as discussed earlier in the paper. Using such a framework will enable practitioners to identify a wide range of metrics that will enable practitioners to capture the requirements and challenges of building security into the software concerning their company and project. We would also like to extend our research by incorporating feedback from more respondents and increasing the number of project teams. Including a greater number of respondents will enable us to reach more concrete answers to our research questions. 

\section*{\uppercase{Acknowledgements}}

This project is in part supported by the California State University San Marcos Professional Development funds.

\bibliographystyle{apalike}
{\small
\bibliography{example}}

\begin{thebibliography}{}

\bibitem[Beznosov and Chess, 2008]{beznosov2008security}
Beznosov, K. and Chess, B. (2008).
\newblock Security for the rest of us: An industry perspective on the
  secure-software challenge.
\newblock {\em IEEE Software}, 25(1):10--12.

\bibitem[Bressan et~al., 2021]{bressan2021variability}
Bressan, L., de~Oliveira, A.~L., Campos, F., and Capilla, R. (2021).
\newblock A variability modeling and transformation approach for
  safety-critical systems.
\newblock In {\em 15th International Working Conference on Variability
  Modelling of Software-Intensive Systems}, pages 1--7.

\bibitem[Forsgren et~al., 2021]{forsgren2021space}
Forsgren, N., Storey, M.-A., Maddila, C., Zimmermann, T., Houck, B., and
  Butler, J. (2021).
\newblock The space of developer productivity: There's more to it than you
  think.
\newblock {\em Queue}, 19(1):20--48.

\bibitem[Fran{\c{c}}a et~al., 2018]{francca2018motivation}
Fran{\c{c}}a, C., Da~Silva, F.~Q., and Sharp, H. (2018).
\newblock Motivation and satisfaction of software engineers.
\newblock {\em IEEE Transactions on Software Engineering}, 46(2):118--140.

\bibitem[G.~Kagombe et~al., 2021]{g2021achieving}
G.~Kagombe, G., Waweru~Mwangi, R., and Muliaro~Wafula, J. (2021).
\newblock Achieving standard software security in agile developments.
\newblock In {\em 2021 The 11th International Conference on Information
  Communication and Management}, pages 24--33.

\bibitem[Graziotin et~al., 2013]{graziotin2013happy}
Graziotin, D., Wang, X., and Abrahamsson, P. (2013).
\newblock Are happy developers more productive?
\newblock In {\em International Conference on Product Focused Software Process
  Improvement}, pages 50--64. Springer.

\bibitem[Imran et~al., 2016]{imran2016web}
Imran, A., Aljawarneh, S., and Sakib, K. (2016).
\newblock Web data amalgamation for security engineering: Digital forensic
  investigation of open source cloud.
\newblock {\em J. UCS}, 22(4):494--520.

\bibitem[Imran et~al., 2013]{imran2013provintsec}
Imran, A., Ul~Gias, A., Rahman, R., and Sakib, K. (2013).
\newblock Provintsec: a provenance cognition blueprint ensuring integrity and
  security for real life open source cloud.
\newblock {\em International Journal of Information Privacy, Security and
  Integrity}, 1(4):360--380.

\bibitem[Ingalsbe et~al., 2008]{ingalsbe2008threat}
Ingalsbe, J.~A., Kunimatsu, L., Baeten, T., and Mead, N.~R. (2008).
\newblock Threat modeling: diving into the deep end.
\newblock {\em IEEE software}, 25(1):28--34.

\bibitem[J{\"u}rjens, 2005]{jurjens2005secure}
J{\"u}rjens, J. (2005).
\newblock {\em Secure systems development with UML}.
\newblock Springer Science \& Business Media.

\bibitem[Khan and Zulkernine, 2008]{4591701}
Khan, M. U.~A. and Zulkernine, M. (2008).
\newblock Quantifying security in secure software development phases.
\newblock In {\em 2008 32nd Annual IEEE International Computer Software and
  Applications Conference}, pages 955--960.

\bibitem[Ko{\l}odziej and Xhafa, 2011]{kolodziej2011meeting}
Ko{\l}odziej, J. and Xhafa, F. (2011).
\newblock Meeting security and user behavior requirements in grid scheduling.
\newblock {\em Simulation Modelling Practice and Theory}, 19(1):213--226.

\bibitem[Kumar et~al., 2008]{kumar2008locking}
Kumar, N., Mohan, K., and Holowczak, R. (2008).
\newblock Locking the door but leaving the computer vulnerable: Factors
  inhibiting home users' adoption of software firewalls.
\newblock {\em Decision Support Systems}, 46(1):254--264.

\bibitem[Lawson and Middleton, 2019]{lawson2019cyber}
Lawson, S. and Middleton, M.~K. (2019).
\newblock Cyber pearl harbor: Analogy, fear, and the framing of cyber security
  threats in the united states, 1991-2016.
\newblock {\em First Monday}.

\bibitem[Rendell, 2019]{rendell2019understanding}
Rendell, D. (2019).
\newblock Understanding the evolution of malware.
\newblock {\em Computer Fraud \& Security}, 2019(1):17--19.

\bibitem[Storey et~al., 2022]{storey2022developers}
Storey, M.-A., Houck, B., and Zimmermann, T. (2022).
\newblock How developers and managers define and trade productivity for
  quality.
\newblock In {\em Proceedings of the 15th International Conference on
  Cooperative and Human Aspects of Software Engineering}, pages 26--35.

\bibitem[Tondel et~al., 2008]{tondel2008security}
Tondel, I.~A., Jaatun, M.~G., and Meland, P.~H. (2008).
\newblock Security requirements for the rest of us: A survey.
\newblock {\em IEEE software}, 25(1):20--27.

\bibitem[Ulsch, 2014]{ulsch2014cyber}
Ulsch, M. (2014).
\newblock {\em Cyber threat!: how to manage the growing risk of cyber attacks}.
\newblock Wiley Online Library.

\bibitem[Zhang et~al., 2022]{zhang2022example}
Zhang, Y., Xiao, Y., Kabir, M. M.~A., Yao, D., and Meng, N. (2022).
\newblock Example-based vulnerability detection and repair in java code.
\newblock In {\em Proceedings of the 30th IEEE/ACM International Conference on
  Program Comprehension}, pages 190--201.

\end{thebibliography}

\end{document}